\documentclass[oupdraft]{bio}
\usepackage[colorlinks=true, urlcolor=citecolor, linkcolor=citecolor, citecolor=citecolor]{hyperref}
\usepackage{url}
\usepackage{mathrsfs}
\usepackage{graphicx,setspace,lscape,longtable}
\usepackage{mathrsfs,amsmath,amsthm,amssymb,color,amstext,mathtools}
\usepackage{natbib,epsfig,graphicx,multirow}
\usepackage{leftidx}
\usepackage{graphicx}
\usepackage{epsfig}
\usepackage{epstopdf,float}
\usepackage{bm}
\usepackage{subfigure}
\usepackage{bbm}
\usepackage{engord}
\usepackage{booktabs}
\usepackage{soul}

\setcitestyle{authoryear,open={(},close={)}} %Citation-related commands

\def\mL{\mathcal L}
\def\argmax{\mbox{argmax}}
\def\argmin{\mbox{argmin}}
\def\mR{\mathbb{R}}
\def\mN{\mathcal N}
\def\D{\mathrm{d}}

\begin{document}

% Title of paper
\title{A Semiparametric Gaussian Mixture Model for Chest CT-based 3D Blood Vessel Reconstruction}

% List of authors, with corresponding author marked by asterisk
\author{QIANHAN ZENG$^1$, JING ZHOU$^{2\ast}$, YING JI$^3$, and HANSHENG WANG$^1$\\[4pt]
% Author addresses
\textit{$^1$Guanghua School of Management, Peking University, China}
\\
\textit{$^2$Center for Applied Statistics, School of Statistics, Renmin University of China, China}
\\
\textit{$^3$ Department of Thoracic Surgery, Beijing Institute of Respiratory Medicine and Beijing Chao-Yang Hospital, Capital Medical University, China}
\\[2pt]
% E-mail address for correspondence
{jing.zhou@ruc.edu.cn}}

% Running headers of paper:
\markboth%
% First field is the short list of authors
{Qianhan Zeng and others}
% Second field is the short title of the paper
{A Semiparametric GMM for Chest CT-based 3D Blood Vessel Reconstruction}

\maketitle

% Add a footnote for the corresponding author if one has been
% identified in the author list
\footnotetext{Corresponding author: jing.zhou@ruc.edu.cn, School of Statistics, Renmin University of China.}

\begin{abstract}
{
{
Computed tomography (CT) has been a powerful diagnostic tool since its emergence in the 1970s.
Using CT data, three-dimensional (3D) structures of human internal organs and tissues, such as blood vessels, can be reconstructed using professional software.
This 3D reconstruction is crucial for surgical operations and can serve as a vivid medical teaching example. 
However, traditional 3D reconstruction heavily relies on manual operations, which are time-consuming, subjective, and require substantial experience.
To address this problem, we develop a novel semiparametric Gaussian mixture model tailored for the 3D reconstruction of blood vessels.
This model extends the classical Gaussian mixture model by enabling nonparametric variations in the component-wise parameters of interest according to voxel positions.
% We theoretically extend the classical Gaussian mixture model by allowing both the component-wise mean and variance to vary according to voxel positions nonparametrically.
% A kernel-based expectation-maximization algorithm is developed to estimate the model and a supporting asymptotic theory is established.
We develop a kernel-based expectation-maximization algorithm for estimating the model parameters, accompanied by a supporting asymptotic theory.
Furthermore, we propose a novel regression method for optimal bandwidth selection.
Compared to the conventional cross-validation-based (CV) method, 
the regression method outperforms the CV method in terms of computational and statistical efficiency.
In application, this methodology facilitates the fully automated reconstruction of 3D blood vessel structures with remarkable accuracy.}
}
{Blood Vessel; Computed Tomography; Gaussian Mixture Model; Nonparametric Kernel Smoothing; TensorFlow; 3D Reconstruction.}
\end{abstract}

\section{Introduction}
\label{sec1}

{
Computed tomography (CT) is an advanced three-dimensional (3D) imaging technology capable of generating detailed, high-resolution 3D images of internal organs and structural intricacies of the human body.
The basic idea of a 3D CT image involves segmenting a lung, as an illustrative example, into several thin pieces.
Each piece is then meticulously scanned using a CT machine, yielding detailed, high-resolution grayscale images.
% This leads to detailed, high-resolution grayscale images. 
Subsequently, these sliced pieces of the lung can be examined individually.
Although examining each CT image slice individually offers practical utility, this approach fails to harness the benefits of 3D technology. 
For example, valuable information embedded within the 3D structure may be overlooked.}

{
In surgical practices, such as lobectomy, wedge resection, or segmentectomy, it is critically important for surgeons to reconstruct blood vessels in three dimensions before surgery. 
This practice enables surgeons to clearly visualize the blood vessels' anatomical structure.
This greatly reduces the risk of accidental or missed blood vessel ruptures during surgery \citep{Hagiwara2014}. 
Furthermore, massive hemorrhages can be avoided \citep{TLCRJi}.
In addition, the 3D reconstructions of blood vessels offer invaluable educational benefits to novice surgeons by facilitating a comprehensive understanding of the lung anatomy and the shape distribution of blood vessels.
However, current standard imaging technology does not yet enable surgeons to perform three-dimensional reconstructions of blood vessels in a fully automated manner.
That is, at present, the 3D reconstruction of blood vessels requires substantial expertise.
For instance, the widely utilized RadiAnt software (\url{https://www.radiantviewer.com/}), recognized for its efficiency and user-friendly interface in viewing CT images, still requires significant manual efforts for such complex reconstructions.
On average, it takes approximately ten minutes for a very experienced surgeon to reconstruct 3D images of blood vessels from raw CT images.
To this end, a sequence of sophisticated but critical, fine-tuned decisions must be made correctly. 
This could be a challenging task for less experienced surgeons.
Consequently, surgical efficiency can be greatly affected. 
Therefore, developing an automatic method for reconstructing 3D blood vessels from raw CT images is of great interest.
}

{To solve this problem, we propose a novel semiparametric Gaussian mixture model (GMM) for CT-based 3D blood vessel reconstruction.
Our method is inspired by the empirical observation that different tissues of different densities manifest different gray intensities in CT images.
Even within the same type of tissue, gray intensities may exhibit slight fluctuations according to their different positions within the human body.
Consequently, CT values corresponding to different organ tissues are expected to manifest different means and variances, while those related to the same organ tissue are anticipated to have similar means and variances. 
This seems to be an ideal situation for applying mixture models for unsupervised clustering analysis.
% The objective here is to cluster voxels associated with the same organ in the same class in a fully automatic manner. 
In this regard, the classical GMM \citep{mixture} appears to be a natural choice. }

{In a GMM, we assume that each type of meaningful human organ tissue (e.g., blood vessels) or the image background is represented by a distinct Gaussian mixture component.}
{Each mixture component is assumed to be a parametric Gaussian model with globally constant mean and variance such that the parameters do not flexibly vary by 3D positions in the CT space.}
{To some extent, this assumption imitates the common practice of standard CT imaging software (e.g., RadiAnt), where only two globally constant, fine-tuned parameters, { such as window width ([WW]) and window level ([WL]),} are allowed. }
% To gain an intuitive understanding, we present in Figure \ref{CT-2} the same CT slice under three different fine-tuned parameters (e.g., WL and WW) using RadiAnt software. 
% These are (from left to right) the bone window (WL: 300, WW: 1500) to demonstrate bone tissue, the chest window (WL: 40, WW: 400) to demonstrate mediastinum tissue, and the lung window (WL: -400, WW: 1500) to demonstrate lung tissue.
{Although this simple practice with globally constant fine-tuned parameters is practically useful, it is not fully satisfactory. 
This is because, even for the same human organ (e.g., blood vessels), the CT density varies according to voxel positions within the 3D CT space.
Furthermore, we observe that the globally constant mean ([WL]) and constant standard deviation ([WW]) are inadequate to accommodate these variations. 
Consequently, the accurate reconstruction of blood vessels in a 3D image remains elusive.
Thus, extending the classical GMM to consider this variation becomes a key issue.}

{To solve this problem, we extend the classical GMM by allowing the parameters of interest, including the mean, variance, and class prior probability functions, to vary nonparametrically.
This variation is due to the different voxel positions within the 3D CT space.
We call this a semiparametric method because it integrates both a parametric component, represented by the GMM, and a nonparametric component, represented by the nonparametric mean, variance, and class prior probability functions.
To estimate the parameters of interest for a given voxel position, a kernel-based expectation-maximization (KEM) algorithm is developed.
Our extensive numerical experiments suggest that this method is effective.
To facilitate computational efficiency, the entire algorithm is presented in a tensor format to efficiently execute the KEM algorithm on a GPU device. 
This design allows for the full utilization of the GPU's parallel computing capabilities. 
Once the parameters of interest are estimated accurately, the posterior probability can be computed for the given voxel position associated with the blood vessel category.
This step is crucial for accurately identifying and reconstructing blood vessels within the 3D CT space.}

{Remarkably, our 3D reconstruction method is very different from the traditional practice commonly used by CT imaging software (e.g., RadiAnt).
In commercial software, raw CT densities are used to reconstruct blood vessels three-dimensionally.
However, as noted, this method results in 3D blood vessel images that lack accuracy.
This is mainly because blood vessels from different voxel positions often have different CT intensity levels, which cannot be fully captured by globally constant fine-tuned parameters.
By contrast, our method reconstructs 3D blood vessels using locally computed posterior probabilities.
By examining CT intensities locally, we can obtain a much clearer vascular anatomy of the blood vessels.
This makes the task of classifying each voxel into the right organ considerably easier.
The locally discovered blood vessels are then represented by their posterior probabilities.
By employing these posterior probabilities as the image intensities, we find that the expression levels of blood vessels from different voxel positions within the 3D CT space are more comparable.
This significantly improves the reconstruction accuracy of the blood vessels.}

{
To summarize, we make two important contributions in this work.
Firstly, we contribute to surgical practice by proposing a fully automated method for 3D blood vessel reconstruction with significantly improved accuracy.
Second, our study enriches the statistical theory of the classical GMM by extending it from a parametric model to a semiparametric one.
The remainder of this paper is organized as follows.
Section 2 introduces the KEM algorithm and elucidates the primary theoretical properties of the proposed method. 
Section 3 presents extensive numerical studies on the KEM method. 
Section 4 concludes this article with a comprehensive discussion of our findings. 
The technical details are included in the Supplementary Materials.
}

\section{METHODOLOGY}
\label{sec2}

\subsection{Model and Notations}

Let $(Y_{i}, X_{i})$ be the observation collected from the $i$th subject, with $Y_{i}\in\mR^{1}$ being the univariate response of interest and $X_{i}\in \mathbb{D} = [0,1]^3$ being the associated voxel position in a three-dimensional Euclidean space. 
We assume that $X_i$ is uniformly distributed on $\mathbb{D}$. 
Furthermore, we assume that for each $i$ a latent class label $Z_{i} \in \{1,\cdots, M\}$, where $M$ is the total number of classes.
We then assume that $P(Z_i = m|X_i) = \pi_m(X_i)$ for $1 \leq m \leq M$.
Clearly, we should have $\sum_{m=1}^M \pi_m(x) = 1$ for any $x \in \mathbb{D}$.
{This suggests that we can define $\pi_M(x) = 1 - \sum_{m=1}^{M-1} \pi_m(x)$ for any $x \in \mathbb{D}$ throughout this article.}
Conditional on $Z_i = m$ and $X_i$, we assume that $Y_i$ is normally distributed with a mean $\mu_m(X_i)$ and variance $\sigma_m^2(X_i)$. 
Here, we assume that class prior probability function $\pi_m(x)$, mean function $\mu_m(x)$, and variance function $\sigma_m^2(x)$ are all smooth functions in $x \in \mathbb{D}$ and vary between classes. 
Next, we consider how to consistently estimate them.

We collect the observed voxel positions and CT values into $\mathbb{X} = \lbrace X_i: 1 \leq i \leq N \rbrace$ and $\mathbb{Y} = \lbrace Y_i: 1 \leq i \leq N \rbrace$, respectively.
Furthermore, we collect the unobserved latent class labels into $\mathbb{Z} = \lbrace Z_i: 1 \leq i \leq N \rbrace$.
For a given voxel position $x$, define $\theta(x) = \{\pi^\top(x), \mu^\top(x), \sigma^\top(x)\}^\top \in \mathbb{R}^{3M-1}$, where $\pi(x) = \{\pi_1(x), \cdots, \pi_{M-1}(x)\}^\top \in \mathbb{R}^{M-1}$, $\mu(x) = \{\mu_1(x), \cdots, \mu_M(x)\}^\top \in \mathbb{R}^{M}$, and $\sigma(x) = \{\sigma_1(x), \cdots, \sigma_M(x)\}^\top \in \mathbb{R}^{M}$. 
Define $\Theta = \{\theta(x): x \in \mathbb{D}\}$. 
To estimate $\theta(x)$, we develop a novel KEM method. 
We begin with a highly simplified case with $\pi_m(x) = \pi_m$, $\mu_m(x) = \mu_m$, and $\sigma_m^2(x) = \sigma_m^2$ for some constants $\pi_m>0$, $\mu_m$, and $0 < \sigma_m < +\infty$. 
We then have the log-likelihood function for an interior point $x$ in $\mathbb{D}$ as
\begin{gather}
    \mL^*\!\Big\{\theta(x)\Big\} = \ln \left\{\prod_{i=1}^N f(X_i, Y_i|\Theta)\right\} = \sum_{i=1}^N \ln \left\{ \sum_{m=1}^M  f(Y_i|m, X_i, \Theta) P(Z_i = m|X_i, \Theta) f(X_i|\Theta)  \right\} \nonumber \\
    = \sum_{i=1}^N \ln\left[\sum_{m=1}^M \phi\!\left\{\frac{Y_i - \mu_m(X_i)}{\sigma_m(X_i)}\right\} \times \left\{ \frac{\pi_m(X_i)}{\sigma_m(X_i)} \right\} \right] \nonumber \\
    = \sum_{i=1}^N \ln\left\{\sum_{m=1}^M \phi\!\left(\frac{Y_i - \mu_m}{\sigma_m}\right) \times \left( \frac{\pi_m}{\sigma_m} \right) \right\}, \label{log-likelihood}
\end{gather}where $f(x, y|\Theta)$ stands for the joint probability density function of $(X_i, Y_i)$ evaluated at $(X_i, Y_i) = (x, y)$; and $f(y|m, X_i, \Theta)$ is the marginal probability density function of $Y_i$ evaluated at $Y_i=y$ conditional on $Z_i = m$ and $X_i$. 
Moreover, $f(x|\Theta)$ is the probability density function of $X_i$ evaluated at $X_i=x$. 
Since $X_i$ is assumed to be uniformly generated on $\mathbb{D}$, we have $f(x|\Theta) = 1$ for $x \in \mathbb{D}$ and $f(x|\Theta) = 0$ for $x \not\in \mathbb{D}$.
Moreover, the function $\phi(y) = \exp(-y^2/2) / \sqrt{2 \pi}$ is the probability density function of a standard normal random variable.

Nevertheless, class prior probability function $\pi_m(x)$, mean function $\mu_m(x)$, and variance function $\sigma_m^2(x)$ in our case are not constant. 
Instead, they should vary {in response to} the value of $x$ (i.e., the voxel position within the 3D CT space) in a fully nonparametric manner.
However, for an arbitrarily given $x$ position, we should expect $\pi_m(x) > 0$, $\mu_m(x)$, and $0 < \sigma_m(x) < +\infty$ to be locally approximately constant. 
This is indeed a reasonable assumption as long as these functions are sufficiently smooth with continuous second-order derivatives. 
{ Thus, according to Proposition 3.10 of \cite{lang2012real}, we then know that the second-order derivatives of $\pi_m(x)$, $\mu_m(x)$, and $\sigma_m(x)$ are uniformly upper bounded from infinity on the compact set $\mathbb{D}$ for any $1 \leq m \leq M$.
Similarly, we know that $\sigma_m(x)$ is uniformly lower bounded on $\mathbb{D}$ by a constant $\sigma_{\min}$ for $1 \leq m \leq M$.}
% Thus, we borrow the idea of local constant estimator \citep{Nadaraya_1965, watson_1964} and local maximum likelihood estimation \citep{fanlocal_1998}, then  propose the following locally weighted log-likelihood function:
Therefore, we draw inspiration from the concept of the local constant estimator \citep{Nadaraya_1965, watson_1964} and local maximum likelihood estimation \citep{fanlocal_1998} and introduce the subsequent locally weighted log-likelihood function {based on the observed data $(\mathbb{X}, \mathbb{Y})$}:
\begin{gather}
    \mL_x(\theta) = \sum_{i=1}^N \ln\left[\sum_{m=1}^M \phi\!\left(\frac{Y_i - \mu_m}{\sigma_m}\right) \times \left(\frac{\pi_m}{\sigma_m}\right) \right] \mathbb{K}\!\left(\frac{X_i - x}{h}\right), \label{kernel weighted log-likelihood}
\end{gather}
where $x=(x_1, x_2, x_3)^\top$ stands for a fixed interior point in $\mathbb{D}$. %\citep{Fan_1996, nonparametric_2007} 
Note that $\theta = (\pi^\top, \mu^\top, \sigma^\top)^\top \in \mathbb{R}^{3M-1}$ is a set of working parameters with $\pi = (\pi_1, \cdots, \pi_{M-1})^\top \in \mathbb{R}^{M-1}$, $\mu = (\mu_1, \cdots, \mu_M)^\top$ $\in \mathbb{R}^M$, and $\sigma = (\sigma_1, \cdots, \sigma_M)^\top \in \mathbb{R}^M$.
{For class $M$, we set $\pi_M = 1 - \sum_{m=1}^{M-1} \pi_m$.
}
Moreover, the three-dimensional kernel function, $\mathbb{K}(\cdot)$, is assumed to be $\mathbb{K}(x) = \mathcal{K}(x_1) \mathcal{K}(x_2) \mathcal{K}(x_3)$, where $\mathcal{K}(t)$ with $t \in \mathbb{R}^{1}$ is a continuous probability density function symmetric about 0. 
Here, $h > 0$ is the associated bandwidth. 
What distinguishes \eqref{kernel weighted log-likelihood} from the classical log-likelihood function \eqref{log-likelihood} is that information on the peer positions of $x$ is also blended in for local estimations. 
For convenience, we refer to $\mL_x(\theta)$ as a locally weighted log-likelihood function. 
Then, the local maximum likelihood estimators can be defined as $\hat{\theta}(x) = \mathop{\argmax}\limits_{\theta} \mL_x(\theta)$.
For convenience, we refer to $\hat{\theta}(x)$ as the kernel maximum likelihood estimators (KMLE). 
We next consider how to optimize $\mL_x(\theta)$ so that $\hat{\theta}(x)$ can be computed.

\subsection{The KEM Algorithm}

{Recall that} the locally weighted log-likelihood function based on the observed data $(\mathbb{X}, \mathbb{Y})$ is already defined in \eqref{kernel weighted log-likelihood}.
{Ever since \cite{dempster1977maximum}, a typical way to optimize \eqref{kernel weighted log-likelihood} is to develop an EM algorithm based on the so-called complete log-likelihood function.
That is the log-likelihood function obtained by assuming that the latent class label $Z_i$ is observed.}
Thus, if we have access to the complete data $(\mathbb{X}, \mathbb{Y}, \mathbb{Z})$, we can define a complete log-likelihood function for the given voxel position $x$ as follows:
\begin{gather}
    \mathcal{Q}_x(\theta) = \sum_{i=1}^N \sum_{m=1}^M I(Z_i = m) \ln \left[ \phi\!\left\{\frac{Y_i - \mu_m(x)}{\sigma_m(x)}\right\} \times \left\{ \frac{\pi_m(x)}{\sigma_m(x)} \right\} \right] \mathbb{K}\!\left(\frac{X_i - x}{h}\right), \label{equation: complete log-likelihood}
\end{gather}
where $I(Z_i = m)$ is an indicator function.
In theory, any location of interest can be taken as $x$.
In practice, a location of interest is often taken at the recorded voxel positions of CT data.
The objective here is to consistently estimate the nonparametric functions $\pi_m(x)$, $\mu_m(x)$, and $\sigma_m(x)$ for $1 \leq m \leq M$. 
We start with a set of initial estimators as $\hat{\pi}_m^{(0)}(x) = 1/M$ and $\hat{\mu}_m^{(0)}(x) = \hat{\mu}_m$, where $\hat{\mu}_m$ is an initial estimator obtained by (for example) a standard $k$-means algorithm \citep{kmeans_1967}. 
Furthermore, we set $\hat{\sigma}_m^{(0)}(x) = \sigma^{(0)}$, where $\sigma^{(0)}$ can be some pre-specified constant.
We write $\hat{\pi}_m^{(t)}(x)$, $\hat{\mu}_m^{(t)}(x)$, and $\hat{\sigma}_m^{(t)}(x)$ as the estimators obtained in the $t$th step. 

\noindent\textbf{E step.}
Based on the current estimate $\hat\theta^{(t)}$, we next take expectations on \eqref{equation: complete log-likelihood} conditional on the observed data $(\mathbb{X}, \mathbb{Y})$ and the current estimate $\hat\theta^{(t)}$ as follows:
\begin{gather}
    E \Big\{\mathcal{Q}_x(\theta) \Big|\mathbb{X}, \mathbb{Y}, \hat\theta^{(t)} \Big\} = \sum_{i=1}^N \sum_{m=1}^M \hat\pi_{im}^{(t)}(X_i) \ln\!\left[ \phi\!\left\{\frac{Y_i - \mu_m(x)}{\sigma_m(x)}\right\} \times \left\{ \frac{\pi_m(x)}{\sigma_m(x)} \right\} \right] \mathbb{K}\!\left(\frac{X_i - x}{h}\right), \label{equation: E Step}
\end{gather}
where $\hat \pi_{im}^{(t)}(X_i) = P(Z_i = m|X_i, Y_i, \hat\theta^{(t)})$.
Specifically, $\hat \pi_{im}^{(t)}(X_i)$ can be computed as follows:
\begin{gather}
    \hat \pi_{im}^{(t)}(X_i) = P(Z_i = m|X_i, Y_i, \hat\theta^{(t)}) = \frac{P(Y_i|X_i, Z_i = m, \hat\theta^{(t)}) P(Z_i = m|X_i, \hat\theta^{(t)})}{\sum_{m=1}^M P(Y_i|X_i, Z_i = m, \hat\theta^{(t)}) P(Z_i = m|X_i, \hat\theta^{(t)})} \nonumber\\
    = \phi\! \left\{ \frac{X_i - \hat\mu_m^{(t)}(X_i)}{\hat\sigma_m^{(t)}(X_i)} \right\} \times \left\{ \frac{\hat\pi_m^{(t)}(X_i)}{\hat\sigma_m^{(t)}(X_i)} \right\} \bigg/ \left[ \sum_{m=1}^M \phi\! \left\{ \frac{X_i - \hat\mu_m^{(t)}(X_i)}{\hat\sigma_m^{(t)}(X_i)} \right\} \times \left\{ \frac{\hat\pi_m^{(t)}(X_i)}{\hat\sigma_m^{(t)}(X_i)} \right\} \right]. \label{E step: pi_im}
\end{gather}
\textbf{M Step.} 
In this step, we maximize \eqref{equation: complete log-likelihood} and obtain a new estimate $\hat\theta^{(t+1)}$ in the $(t+1)$th step.
Note that we have $\sum_{m=1}^M \pi_m(x) = 1$ for any $x\in \mathbb{D}$.
Based on the Lagrange method, we can define the Lagrangian function as $\mathcal{Q} = E\lbrace\mathcal{Q}_x(\theta)|\mathbb{X}, \mathbb{Y}, \hat\theta^{(t)}\rbrace + \lambda(\sum_{m=1}^M \pi_m - 1)$, where $\lambda$ is an additional parameter introduced by the Lagrange multiplier method.
After maximizing $\mathcal{Q}$, we can update the parameters in the $(t+1)$th step by
\begin{gather}
    % M step
    \hat{\pi}_m^{(t+1)}(x) = \sum_{i=1}^N \hat{\pi}_{im}^{(t)}(X_i) \mathbb{K}\!\left(\frac{X_i - x}{h}\right) \Big/ \sum_{i=1}^N \mathbb{K}\!\left(\frac{X_i - x}{h}\right), \label{M step: pi_m} \\
    \hat{\mu}_m^{(t+1)}(x) = \sum_{i=1}^N \hat{\pi}_{im}^{(t)}(X_i) \mathbb{K}\!\left(\frac{X_i - x}{h}\right) Y_i \Big/ \sum_{i=1}^N \hat{\pi}_{im}^{(t)}(X_i) \mathbb{K}\!\left(\frac{X_i - x}{h}\right), \label{M step: mu_m} \\
    \hat{\sigma}_m^{(t+1)}(x) = \left[\sum_{i=1}^N \Big\{Y_i - \hat{\mu}_m^{(t+1)}(X_i)\Big\}^2 \mathbb{K}\!\left(\frac{X_i - x}{h}\right) \Big/ \sum_{i=1}^N \hat{\pi}_{im}^{(t)}(X_i) \mathbb{K}\!\left(\frac{X_i - x}{h}\right) \right]^{\frac{1}{2}}. \label{M step: sigma_m}
\end{gather}
For convenience, we refer to \eqref{E step: pi_im}--\eqref{M step: sigma_m} as a KEM algorithm, which should be iteratively executed until convergence. 
% By the time of convergence, we obtain the final estimators, $\hat\theta(x)$.
Our extensive numerical experiments suggest that the algorithm works very well.

\subsection{Asymptotic Properties}

Next, we study the asymptotic properties of KMLE $\hat\theta(x)$. First, we define some notations. 
Recall that $\mathcal{L}_x(\theta)$ is the locally weighted log-likelihood function defined in \eqref{kernel weighted log-likelihood}. 
We define $\nu_{k, m} = \int t^k \mathcal{K}^m(t) \D t$ for $0 \leq k \leq 2$ and $1 \leq m \leq 2$. 
The following technical conditions are then required:
\begin{itemize}
\item[(C1)] (\textit{Kernel Function}) 
% Assume that $|\nu_{k,m}| < +\infty$ for $0 \leq k \leq 2$ and $1 \leq m \leq 2$.
Assume $|\nu_{k, m}| < +\infty$ for $1 \leq m \leq 2$ and $0 \leq k \leq 2+\delta$ with some $\delta > 0$.
\item[(C2)] (\textit{Bandwidth and Sample Size}) Assume $h = C_h N^{-1/7}$ for some constant $C_h > 0$.
\end{itemize}
As we mentioned before, we assume that kernel function $\mathbb{K}(t)$ is the product of three univariate kernel density functions symmetric about 0. 
By Condition (C1), each univariate kernel density function should be a well-behaved probability density function with various finite moments. 
By Condition (C2), we require the bandwidth $h$ to converge to zero at the speed of $N^{-1/7}$ as the sample size, $N$, goes to infinity. 
As a consequence, the locally effective sample size, denoted by $Nh^3$, diverges toward infinity as $N \rightarrow +\infty$.
In the meanwhile, the bandwidth $h$ is well constructed so that the resulting estimation bias should not be too large. 
Both Conditions (C1) and (C2) are fairly standard conditions, which have been widely used in the literature \citep{Fan_1996, pagan_ullah_1999, nonparametric_2007, silverman_1986}.

% Note that $\mathcal{L}_x\{\theta(x)\}$ with respect to $\theta(x)$ is an over-parameterized log-likelihood function because $\sum_{m=1}^M \pi_m(x) = 1$ for any $x \in \mathbb{D}$. 
% Computationally, this is not an issue because the Lagrange multiplier method can be developed to derive an EM algorithm. 
% However, it is inconvenient to study the asymptotic properties of estimators obtained by optimizing $\mathcal{L}_x\{\theta(x)\}$.
% A more convenient approach is to redefine this log-likelihood function with respect to $\theta(x)=\{\pi^{*\top}(x), \mu^\top(x), \sigma^\top(x) \} \in \mathbb{R}^{3M-1}$, where $\pi^*(x) = \{\pi_1(x), \cdots, \pi_{M-1}(x)\}^\top \in \mathbb{R}^{M-1}$. 
% Thus, the log-likelihood function $\mathcal{L}_x\{\theta(x)\}$ becomes $\mathcal{Q}_x\{\theta(x)\}$ in terms of $\theta(x)$.
We define $I\{\theta(x)\} = \int \partial \ln f(x, y|\Theta) / \partial \theta(x) \times \partial \ln f(x, y|\Theta) /{\partial \theta(x)^\top} \times f(x, y|\Theta) \D y$.
Denote the Euclidean norm as $\|z\| = \sqrt{z^\top z}$ for any vector $z$.
Then, we have the following Theorem \ref{theorem1}, with the technical details provided in Appendix A and Appendix B of the Supplementary Materials.

\begin{theorem} \label{theorem1}
Assume that both Conditions (C1) and (C2) are satisfied, then (i) there exists a local maximum likelihood estimator $\hat\theta(x)$ such that $\|\hat\theta(x) - \theta(x)\| = O_p(1 / \sqrt{Nh^3})$; and (ii) $\sqrt{Nh^3} \left[ \hat\theta(x) - \theta(x) -  \nu_{2,1} h^2 I^{-1}\big\{\theta(x)\big\} g(x)/{2}\right] \stackrel{d}{\rightarrow} \mathcal{N}\left[\mathbf{0}, \nu_{0, 2}^3 I^{-1}\big\{\theta(x)\big\} \right]$, where $g(x)$ is given by
$g(x)=\int \lbrace{\partial \ln f(x, y|\Theta)}/{\partial \theta(x)}\rbrace {\rm tr}\{ {\partial^2 f(x, y|\Theta)}/{\partial x \partial x^\top} \} \D y.$
\end{theorem}

\subsection{Bandwidth Selection through Cross Validation} \label{section: CV}

To ensure the asymptotic properties of the KMLE, we must carefully select the optimal bandwidth $h$. From Condition (C2), we know that the optimal bandwidth should satisfy $h = C_h N^{-1/7}$. 
% By Theorem \ref{theorem1}, we further know that the asymptotic bias of $\hat\theta(x)$ is given by $C_h^2 N^{-2/7} I^{-1}\{\theta(x)\}g(x)/2$ and the asymptotic variance is given by $C_h^{-3} \nu_{0, 2}^3 I^{-1}\{\theta(x)\} N^{-4/7}$. 
The question is how to make an optimal choice for $C_h$.
To do so, an appropriately defined optimality criterion is required. 
One natural criterion could be out-of-sample forecasting accuracy. 
Let $(X^*, Y^*)$ be an independent copy of $(X_i, Y_i)$. 
Recall that $\hat\theta(x)=\{\hat\pi_1(x), \cdots, \hat\pi_{M-1}(x), \hat\mu_1(x), \cdots,$  $ \hat\mu_M(x), \hat\sigma_1(x), \cdots, \hat\sigma_M(x)\}^\top$ are the estimators obtained from data $\{(X_i, Y_i): 1 \leq i \leq N\}$. Note that $E(Y^*|X^*, \Theta)=\sum_{m=1}^M \pi_m(X^*) \mu_m(X^*)$. Therefore, a natural prediction for $Y^*$ can be constructed as $\hat{Y}^* = \sum_{m=1}^M \hat\pi_m(X^*) \hat\mu_m(X^*)$. Its square prediction error (SPE) can then be evaluated as $E(Y^* - \hat{Y}^*)^2$. Using a standard Taylor expansion-type argument, we can obtain an analytical formula for SPE using the following theorem:

\begin{theorem} \label{theorem2}
Assume that both Conditions (C1) and (C2) are satisfied, we then have
$$
E\Big( \hat{Y}^* - Y^* \Big)^2 = \Big( \sigma_y^2 + \frac{1}{4} C_h^4 N^{-4/7} C_1 + C_h^{-3} N^{-4/7} C_2 \Big) \{1 + o(1)\},
$$
where $\sigma_y^2 = \int_\mathbb{D} \sum_{m=1}^M \pi_m(x) \{ \sigma_m^2(x) + \mu_m^2(x) \} - \mu^2(x) \D x$, $\mu(x) = \sum_{m=1}^M \pi_m(x) \mu_m(x)$, $C_1 = \int_\mathbb{D} [\nu_{2,1} {\dot{\varphi}\{\theta(x)\}^\top} {I^{-1}\{\theta(x)\}} {g(x)} ]^2 \D x$, and $C_2 = \int_\mathbb{D} \nu_{0, 2}^3 \dot{\varphi} \{\theta(x) \}^\top I^{-1}\{\theta(x) \} \dot{\varphi} \{\theta(x) \} \D x$. 
Here, $\varphi \{\theta(x)\} = \sum_{m=1}^{M-1} \pi_m(x) \{ \mu_m(x) - \mu_M(x) \} + \mu_M(x)$ and $\dot{\varphi}\{\theta(x)\} = \partial \varphi \{\theta(x)\} / \partial \theta(x)$.
\end{theorem}

The technical details of Theorem \ref{theorem2} are provided in Appendix C in the Supplementary Materials. 
By optimizing the leading term of $E(\hat{Y}^* - Y^*)^2$ from Theorem \ref{theorem2}, we know that the optimal bandwidth constant is given by $C_h^* = (3C_2 / C_1)^{1/7}$. 
However, both the critical constants, $C_1$ and $C_2$, are extremely difficult to compute without explicit assumptions on the concrete forms of $\pi(x)$, $\mu(x)$, and $\sigma(x)$ with respect to $x$.
To solve this problem, a cross-validation (CV) type method is used to compute the optimal bandwidth. To this end, we need to randomly partition all voxel positions into two parts. The first part contains approximately 80\% of the total voxels used for training. The remaining 20\% is used for testing. 
For convenience, the indices of the voxels in the training and testing dataset are collected as $\mathcal{I}_0$ and $\mathcal{I}_1$, respectively. Next, for an arbitrary testing sample, $i \in \mathcal{I}_1$, we have $E(Y_i|X_i) = \sum_{m=1}^M \pi_m(X_i) \mu_m(X_i)$. 
Therefore, we are inspired to predict the $Y_i$ value using $\hat{Y}_i = \sum_{m=1}^M \hat{\pi}_m(X_i) \hat{\mu}_m(X_i)$, where the estimates $\hat{\pi}_m(X_i)$ and $\hat{\mu}_m(X_i)$ are computed based on the training data with a given bandwidth constant. Let $\mathbb{C}_{\text{CV}}=\{C_h^{(g)}: 1 \leq g \leq G_{\text{CV}}\}$ be a set of tentatively selected pilot-bandwidth constants, where $G_{\text{CV}}$ is the total number of pilot-bandwidth constants. 
Therefore, an estimator for SPE can be constructed as $\hat{\mathscr{L}}_{\text{SPE}}(C_h^{(g)}) = |\mathcal{I}_1|^{-1} \sum_{i \in \mathcal{I}_1} ( Y_i - \hat{Y}_i )^2$.
A CV-based estimator for the optimal bandwidth constant can then be defined as $\hat{C}_h^{\text{CV}} = \mathop{\argmin}\limits_{C_h^{(g)} \in \mathbb{C}_{\text{CV}}} \hat{\mathscr{L}}_{\text{SPE}}(C_h^{(g)})$; that is, the bandwidth constant in $\mathbb{C}_{\text{CV}}$ with the lowest SPE value is chosen. Consequently, a CV-based estimator for the optimal bandwidth is given by $\hat{h}^{\text{CV}} = \hat{C}_h^{\text{CV}} \times N^{-1/7}$.

\subsection{Bandwidth Selection through Regression} \label{section: REG}

Even though the above CV idea is intuitive, its practical computation is expensive.
This is because, for an accurate estimation of the optimal bandwidth constant $C_h^*$, we typically need to evaluate a large number of pilot-bandwidth constants. 
To reduce the computation cost, we develop a novel regression method as follows.
From Theorem \ref{theorem2}, we know that $E\{\hat{\mathscr{L}}_{\text{SPE}}(C_h)\} \approx \sigma_y^2 + (C_h^4 C_1 / 4 + C_h^{-3} C_2) N^{-4/7}$.
Note that $E\{\hat{\mathscr{L}}_{\text{SPE}}(C_h)\}$ is approximately a linear function in both $C_1$ and $C_2$. 
The corresponding weights are given by $C_h^4 N^{-4/7} / 4$ and $C_h^{-3} N^{-4/7}$, where $N$ is the given total sample size, and $C_h$ is the tentatively selected pilot-bandwidth constant. 
This immediately suggests an interesting regression-based method for estimating the two critical constants, $C_1$ and $C_2$, as follows.
We define $\mathbb{C}_\text{REG}=\{C_h^{(g)}: 1 \leq g \leq G_\text{REG}\}$ as a set of carefully selected pilot-bandwidth constants.
Note that $G_\text{REG}$ determines the total number of pilot-bandwidth constants to be tested. 
It must not be too large so that the computation cost can be reduced. 
For example, we fix $G_{\text{CV}} = 25$ and $G_\text{REG} = 5$ in our subsequent numerical studies. 
We define a pseudo response, $\mathbb{Y} = \{ \mathscr{L}_{\text{SPE}}(C_h^{(g)}) - \bar{\mathscr{L}}_{\text{SPE}}: 1 \leq g \leq G_\text{REG}\}^\top \in \mathbb{R}^{G_\text{REG}}$, where $\bar{\mathscr{L}}_{\text{SPE}} = G_\text{REG}^{-1} \sum_{g=1}^{G_\text{REG}} \mathscr{L}_{\text{SPE}}(C_h^{(g)})$. 
We define $X^{(g)} = (N^{-4/7} C_h^{(g)4} / 4, N^{-4/7} /C_h^{(g)3} )^\top \in \mathbb{R}^2$.
We further define a pseudo-design matrix, $\mathbb{X} \in \mathbb{R}^{G_\text{REG}}$, where the $g$th row is $(X^{(g)} - \bar{X})^\top$ and $\bar{X} = G_\text{REG}^{-1} \sum_{g=1}^{G_\text{REG}} X^{(g)}$.
% $\mathbb{X} = \{I_{G_\text{REG}} - \mathbf{1}_{G_\text{REG}} \mathbf{1}_{G_\text{REG}}^\top / {G_\text{REG}}\} (X^{(1)\top}, \cdots, X^{({G_\text{REG}})\top})^\top \in \mathbb{R}^{{G_\text{REG}} \times 2}$, where $X^{(g)} = (N^{-4/7} C_h^{(g)4} / 4, N^{-4/7} /C_h^{(g)3} )^\top$, $I_{G_\text{REG}} \in \mathbb{R}^{{G_\text{REG}} \times {G_\text{REG}}}$ is an identity matrix, and $\mathbf{1}_{G_\text{REG}} = (1, \cdots, 1)^\top \in \mathbb{R}^{G_\text{REG}}$. 
Then, we have an appropriate regression relationship as $\mathbb{Y} = \mathbb{X} \mathcal{C}$, where $\mathcal{C} = (C_1, C_2)^\top \in \mathbb{R}^2$. 
Subsequently, an ordinary least squares estimator is obtained for $\mathcal{C}$ as $\hat{\mathcal{C}} = (\mathbb{X}^\top \mathbb{X})^{-1} (\mathbb{X}^\top \mathbb{Y}) = (\hat{C}_1, \hat{C}_2)^\top$. 
This leads to a regression-based estimator for the optimal bandwidth constant $C_h^*$ as $\hat{C}_h^{\text{REG}}=(3\hat{C}_2/\hat{C}_1)^{1/7}$.
For convenience, we abbreviate this method as the REG method.

\section{NUMERICAL STUDIES}
\label{sec3}

\subsection{The Simulation Model} \label{section: simulation setting}

Extensive simulation studies are conducted to validate the finite sample performance of the proposed KEM method. 
The entire experiment is designed so that the simulation data can mimic real chest CT data as much as possible. Specifically, we first fix a total of $M=3$ mixture components, which represent the background, bone tissue, and lung tissue, respectively. 
%Those three mixture components are arguably the three most important parts, which need to be distinguished from each other by a chest CT for surgical purposes. 
Next, we need to specify the class prior probability function, $\pi_{m}(x)$. To make the experiment as realistic as possible, we use the LIDC-IDRI dataset, which is probably the largest publicly available chest CT database \citep{luna2016}. It consists of 888 human chest CT files from seven medical institutions worldwide. 
After downloading the dataset, we found that six files were damaged and could not be read into memory. Thus, we use the remaining 882 CT files for the subsequent study.

Let $\mathbb{Y} = (Y_{ijk})\in \mR^{d_{x}\times d_{y} \times d_{z}}$ be an arbitrarily selected chest CT data from the LIDC-IDRI dataset. 
According to the definition of the database, we should always have $d_{x}=d_{y}=512$ for every CT scan file. However, the number of slices (i.e., $d_{z}$) may vary across different CT scan files, but it should fall within the range of $d_{z} \in [95, 764]$.
Next, we consider how to specify the values of $\pi_{m}(X_{ijk})$ for every possible voxel position $X_{ijk} = (i/d_x, j/d_y, k/d_z)^\top \in \mathcal{V}$, where $\mathcal{V} \subset [0,1]^3$ contains all the available voxel positions.
Next, we define a class label tensor $\mathbb{Z} = (Z_{ijk}) \in \mR^{d_{x}\times d_{y} \times d_{z}}$ with $Z_{ijk}\in \{1,2,3\}$. Specifically, we define $Z_{ijk}=2$ if its voxel position, $X_{ijk}$, belongs to the lung-mask data provided by the LIDC-IDRI dataset. 
Otherwise, we define $Z_{ijk}=3$ if $Y_{ijk}>400$. 
In this case, the $(i, j, k)$th voxel position should be the bone tissue \citep{bone_2013}. Finally, we define $Z_{ijk}=1$ if the previous two criteria are not met.

\begin{figure}[ht]
    \centering
    {\includegraphics[scale=0.45]{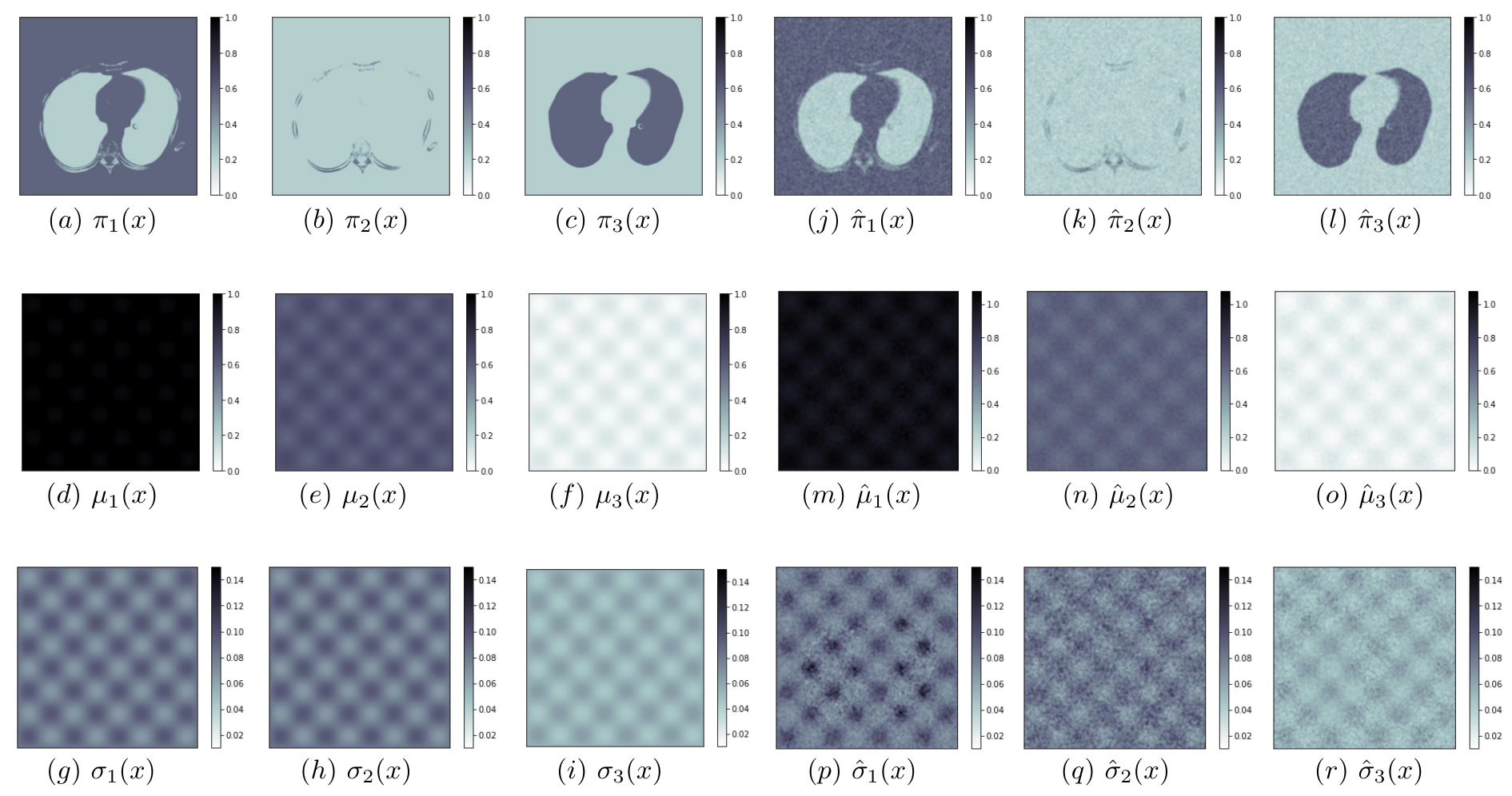}}
    \caption{Graphical displays of $\pi_m(x)$, $\mu_m(x)$, $\sigma_m(x)$, $\hat\pi_m(x)$, $\hat\mu_m(x)$, and $\hat\sigma_m(x)$ for $1 \leq m \leq 3$ based on an arbitrarily selected chest CT slice.}
    \label{pic: parameters}
\end{figure}

For the $(i, j, k)$th voxel position $x = (i/d_x, j/d_y, k/d_z)^\top = (x_{1}, x_{2}, x_{3})^\top \in \mathcal{V}$, we define $x$ for a local neighborhood as $\mN(x) = \big\{(x_1', x_2', x_3')^\top: |(x_1' d_x, x_2' d_y, x_3' d_z)^\top - (x_1 d_x, x_2 d_y, x_3 d_z)^\top|_{\max} < 3 \big\}$, where $|x|_{\max} = \mathop{\max}\limits_{1 \leq j \leq 3} |x_j|$. Then, we define $\pi_{m}(x) = \{ \sum_{x' \in \mN(x)}I(Z_{i'j'k'}=m) \} / |\mN(x)|$, where $x' = (i'/d_x, j'/d_y, k'/d_z)^\top = (x_{1}', x_{2}', x_{3}')^\top \in \mathcal{V}$. 
Next, we redefine $\pi_m(x) = \{\pi_m(x) + 0.6\} / 2.8$ for $1 \leq m \leq M$. By doing so, we can guarantee that $0 < \pi_m(x) < 1$ for every $1 \leq m \leq M$ and $x \in \mathcal{V}$. 
Next, we define $\mu_{m} = \{ \mathop{\sum}\limits_{x \in \mathcal{V}} I(Z_{ijk}=m) Y_{ijk}\} / \{  \mathop{\sum}\limits_{x \in \mathcal{V}} I(Z_{ijk}=m)\}$ and
$\sigma_{m}^2 = \mathop{\sum}\limits_{x \in \mathcal{V}} \{ I(Z_{ijk}=m) (Y_{ijk} - \mu_m)^2\} / \{  \mathop{\sum}\limits_{x \in \mathcal{V}} I(Z_{ijk}=m)  \}$ for $2 \leq m \leq M$. 
The $m=1$ case is set as the background case. 
To differentiate the background from the other classes, $\mu_1$ is set to 1, and $\sigma_1$ is set to $\sigma_3$.
Then, we set $\mu_m(x) = \mu_m + 0.25 \times \sin(8\pi x_1) \times \sin(8\pi x_2) \times \sin(8\pi x_3)$ and $\sigma_{m}(x) = \sigma_m + \sigma_m \times \sin(8\pi x_1) \times \sin(8\pi x_2) \times \sin(8\pi x_3) $.
Thus, we allow the mean function $\mu_m(x)$ and variance function $\sigma_m^2(x)$ to vary within a reasonable range. 
For intuitive understanding, $\pi_m(x)$, $\mu_m(x)$, and $\sigma_m(x)$ with $1 \leq m \leq 3$ generated from an arbitrarily selected CT are graphically displayed in Figure \ref{pic: parameters}(a)--(i). 
Throughout this article, the depth index for the displayed 2D slices is consistently set to 100.
%In accordance with our simulation setup, it becomes apparent, as observed in panels (a)--(c), that while each position holds the potential to manifest any of the three specified categories, panel (a) predominantly portrays the background, panel (b) depicts bone structures, and panel (c) represents the lung mask. In contrast to the category probability, the visualization of the true mean and standard deviation functions (i.e., $\mu_{m}(x)$ and $\sigma_{m}(x)$) showcased in panels (d)--(i), reveals intricate spatial variations. 
For any voxel position $x \in \mathcal{V}$, once $\mu_m(x)$ and $\sigma_m(x)$ are given, a normal random variable can be generated. 
These normal random variables are then combined across different components according to a multinomial distribution with probability $\pi_m(x)$ for class $m$ with $1 \leq m \leq M$. 
This leads to a final response $Y_{ijk}$.

\subsection{Implementation Details}

Although the KEM algorithm developed in equations \eqref{E step: pi_im}--\eqref{M step: sigma_m} is theoretically elegant, its computation is not immediately straightforward. 
If one implements the algorithm faithfully, as equations \eqref{E step: pi_im}--\eqref{M step: sigma_m} suggest, the associated computational cost would be unbearable.
Consider, for example, the computation of the denominator of equation \eqref{M step: pi_m}, which is $\sum_{i=1}^N \mathbb{K}(X_i / h - x/h)$.
The resulting computational complexity is $O(N)$, where $N=|\mathcal{V}|=d_x \times d_y \times d_z$ is the total number of voxel positions. 
Note that this procedure should be performed for each voxel position in $\mathcal{V}$. 
Then, the overall computational cost becomes $O(N^2)$ order.
Consider, for example, a chest CT of size $512 \times 512 \times 201$; then, we have $N = 5.27 \times {10}^{7}$ and $N^2 = 2.78 \times {10}^{15}$. 
Thus, the computational cost is prohibitive, and alleviating the computational burden becomes a critical problem.

To address this problem, we develop a novel solution. 
We specify $\mathcal{K}(t) = \exp (-t^2/2) / \sqrt{2 \pi}$ as the Gaussian kernel and $\mathbb{K}(x) = \mathcal{K}(x_1) \mathcal{K}(x_2) \mathcal{K}(x_3)$. 
Accordingly, the weight assigned to its tail decays toward 0 at an extremely fast rate. 
Therefore, instead of directly using the full kernel weight $\mathbb{K}(x)$, we consider its truncated version as $\mathbb{K}^*(x) = \mathbb{K}(x) I(|(x_1 d_x, x_2 d_y, x_3 d_z)^\top|_{\max} < s)$, where $s > 0$ represents the filter size.
We define $\mathcal{N}_s(x) = \{x'=(x_1', x_2', x_3')^\top \in \mathcal{V}: |(x_1' d_x, x_2' d_y, x_3' d_z)^\top - (x_1 d_x, x_2 d_y, x_3 d_z)^\top|_{\max} < s\}$ as a cubic space locally around $x$ with volume $|\mathcal{N}_s(x)| = s^3$. 
Instead of computing $\sum_{i=1}^N \mathbb{K}(X_i/h - x/h)$, we can compute $\sum_{i=1}^N \mathbb{K}^*(X_i/h-x/h) = \sum_{1 \leq i\leq N}^{X_i \in \mathcal{N}_s(x)} \mathbb{K}(X_i/h-x/h)$. 
Thus, the computational cost for one voxel position can be significantly reduced from $O(N)$ to $O(s^3)$, which is approximately the size of set $\{X_i \in \mathcal{N}_s(x): 1 \leq i \leq N\}$. 
This is typically a much-reduced number compared with $N$. 
Furthermore, this operation can be easily implemented on a GPU device using a standard 3D convolution operation (e.g., the Conv3D function in Keras of TensorFlow). 
This substantially accelerates the computational speed.
To this end, the bandwidth must be appropriately specified according to the filter size of the convolution operations (i.e., $s$).
Ideally, the filter should not be too large. 
Otherwise, the computation cost owing to $|\mathcal{N}_s(x)|$ could be extremely high.
However, the filter size, $s$, cannot be too small compared to bandwidth $h$. 
Otherwise, the probability mass of $\mathbb{K}(x)$, which falls into $\mathcal{N}_s(x)$, could be too small.

For the REG method, we consider the filter size as $s^{(g)} = 2g+1$, where $1 \leq g \leq G_{\text{REG}}$ and $G_{\text{REG}} = 5$. 
Accordingly, we set the bandwidth to $h^{(g)} = s^{(g)} / 512$. 
Thus, only odd numbers are considered, so the filter used in the 3D convolution operations can be symmetric about its central voxel position.
We wish the filter to be as large as possible.
However, as previously discussed, an unnecessarily large filter size would result in prohibitive computational costs. 
Therefore, we set the largest filter size as 11.
For the CV method, following the REG method, we generate $(s^{(g)}, h^{(g)})$ for $1 \leq g \leq G_{\text{REG}}$. 
Subsequently, each $h^{(g)}$ is multiplied by 0.3, 0.4, 0.5, 0.6, or 0.7. 
% Thus, the CV method generates 25 $(s^{(g)}, h^{(g)})$--pairs, with $G_{\text{CV}}=25$.
The bandwidth constant of the two methods is set as $C_h^{(g)} = h^{(g)} \times N^{1/7}$, where $N$ represents the sample size.

\subsection{Bandwidth Selection and Estimation Accuracy}

\begin{figure}
    \centering
    \includegraphics[scale=0.45]{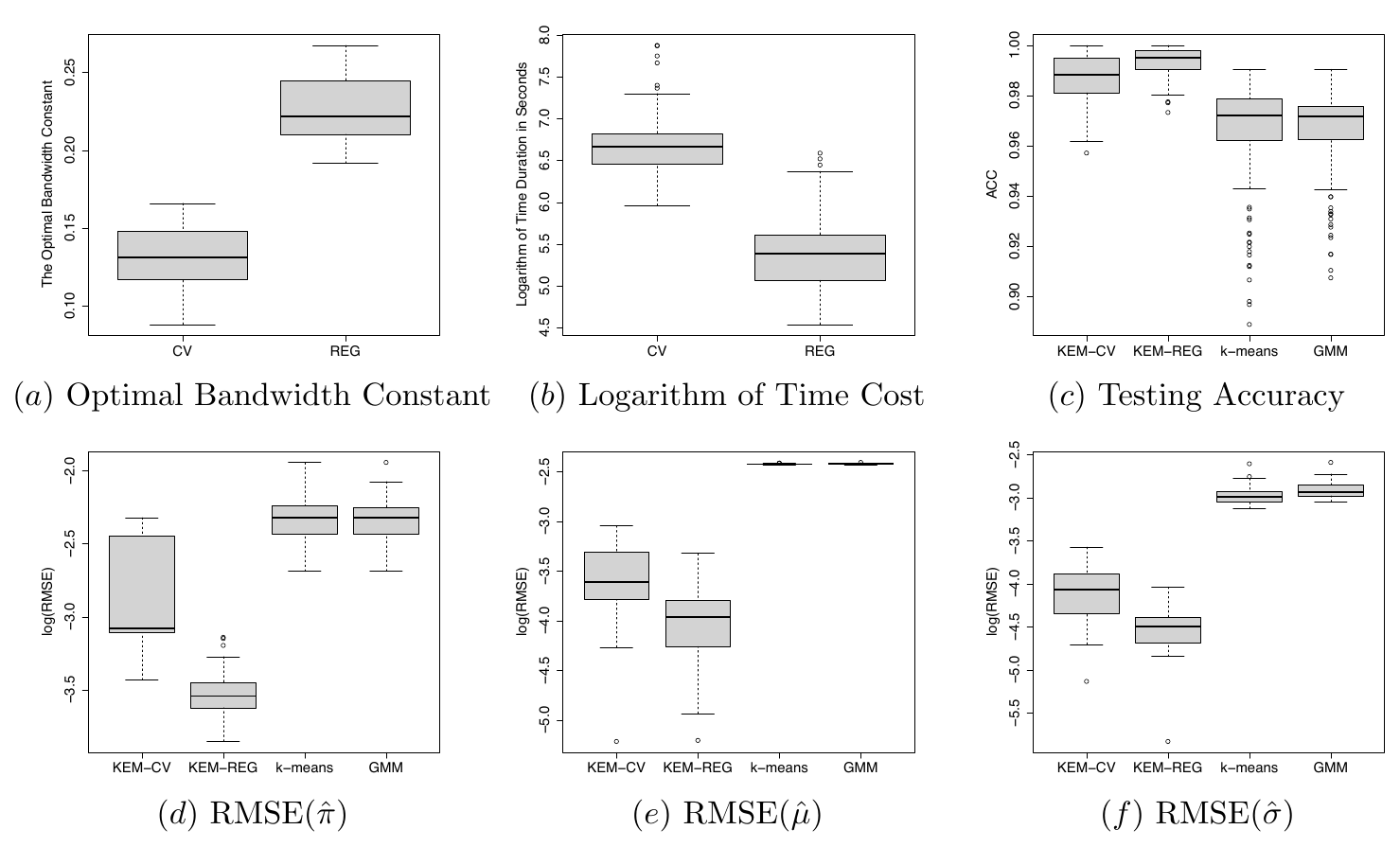}
    % \vskip -0.1 in
    \caption{Panel (a) is the boxplot of the optimal bandwidth constant $C_h$ selected by the CV and REG methods. 
    Panel (b) is the logarithm of the time cost of the two methods in seconds. 
    Panel (c) is the boxplot of the testing accuracy of the KEM under two bandwidth selection methods, the $k$-means method, and the GMM method.
    Panels (d)--(f) represent $\mbox{RMSE}(\hat{\pi})$, $\mbox{RMSE}(\hat{\mu})$, and $\mbox{RMSE}(\hat{\sigma})$ of the concerned methods, respectively.}
    \label{boxplot: bwd mse bosplot}
\end{figure}

With the pre-specified filter sizes, both the CV and REG methods introduced in Sections \ref{section: CV} and \ref{section: REG} can be used to select the optimal bandwidth.
Recall that 80\% of the voxels are used for training and the remaining 20\% are used for testing. 
Truncated Gaussian kernel and GPU-based convolutions are used to speed up the practical computation.
Both the CV and REG methods are used to select the optimal bandwidth constant for each of the 100 randomly selected patients in the LIDC-IDRI dataset.
The optimal bandwidth constants selected by the two methods are then boxplotted in Figure \ref{boxplot: bwd mse bosplot}(a).
We find that the REG method tends to select a larger $C_h$ than the CV method, leading to relatively larger filter sizes for convolution. 
The time cost of the two methods for selecting the optimal bandwidth for each patient is boxplotted in Figure \ref{boxplot: bwd mse bosplot}(b).
% The CV method costs 358.5s to compute the $C_h$ for one patient on average, while the REG method only needs about 139.4s on average. 
Evidently, the REG method has a much lower computational cost than the CV method.

To show the superiority of the proposed KEM method, we also compared it with two traditional methods.
The two traditional methods in concern are, respectively, $k$-means \citep{kmeans_1967} and GMM \citep{mixture}.
We then compare the aforementioned methods in terms of testing accuracy and estimation accuracy.
Testing accuracy reflects the ability to identify the class labels on the testing set, while estimation accuracy reflects the ability to recover the true parameter values.
% The estimation accuracy is also compared based on the optimal bandwidth constants obtained using the CV and REG methods.
When comparing the testing accuracies, we train the model on the training set and evaluate the testing accuracy on the testing set.
See Figure \ref{boxplot: bwd mse bosplot}(c) for the results of the testing accuracy.
The medians of the testing accuracy results for the KEM method with the bandwidth selected by the CV method (KEM-CV), the KEM method with the bandwidth selected by the REG method (KEM-REG), the $k$-means method, and the GMM method are, respectively, 0.9883, 0.9950, 0.9721, and 0.9719.
% We can see that the KEM(REG) method slightly outperforms the KEM(CV) method, which again advocates the usage of the REG method for optimal bandwidth selection.
We can see that both the traditional methods suffer from worse testing accuracy as compared with the proposed KEM methods.
When comparing the estimation accuracies, all voxel positions of the CT data are used. 
We can compute $\hat{\pi}_m(x)$, $\hat{\mu}_m(x)$, and $\hat{\sigma}_m(x)$ using the KEM-CV, KEM-REG, $k$-means and GMM methods for every $1 \leq m \leq M$ and every $x \in \mathcal{V}$. 
% We define $\hat{\pi}(x) = \{ \hat{\pi}_m(x): 1 \leq m \leq M\} \in \mathbb{R}^M$, $\hat{\mu}(x) = \{ \hat{\mu}_m(x): 1 \leq m \leq M\} \in \mathbb{R}^M$, and $\hat{\sigma}(x) = \{ \hat{\sigma}_m(x): 1 \leq m \leq M\} \in \mathbb{R}^M$.
Because this is a simulation study with the true parameter values specified in Section \ref{section: simulation setting}, which are known to us, we can evaluate the accuracy of the resulting estimates using the following RMSE criteria:
\begin{gather}
    \text{RMSE}(\hat{\pi}) = \left[ |\mathcal{V}|^{-1} m^{-1} \sum_{x \in \mathcal{V}} \sum_{m=1}^M \Big\{\hat{\pi}_m(x) - \pi_m(x)\Big\}^2 \right]^{1/2}, \nonumber
    \\
    \text{RMSE}(\hat{\mu}) = \left[ |\mathcal{V}|^{-1} m^{-1} \sum_{x \in \mathcal{V}} \sum_{m=1}^M \Big\{\hat{\mu}_m(x) - \mu_m(x)\Big\}^2 \right]^{1/2}, \nonumber \\
    \text{RMSE}(\hat{\sigma}) = \left[|\mathcal{V}|^{-1} m^{-1} \sum_{x \in \mathcal{V}} \sum_{m=1}^M \Big\{\hat{\sigma}_m(x) - \sigma_m(x)\Big\}^2 \right]^{1/2}. \nonumber
\end{gather}
The RMSE results of the aforementioned four methods are boxplotted in Figures \ref{boxplot: bwd mse bosplot}(d)--(f). 
We find that the KEM method outperforms the $k$-means and GMM methods substantially.
This is not surprising, since the two traditional methods cannot estimate the parameters of interest nonparametrically.
Comparatively speaking, the REG method performs slightly better than the CV method. 
Thus, the median value of the bandwidth constants (i.e., $C_h=0.2217$) chosen by the REG method is then fixed for the subsequent simulation studies.

With the fixed bandwidth constant $C_h=0.2217$, we conduct more comprehensive simulation studies to evaluate the finite sample performance of the estimators. 
These are, respectively, $\hat{\pi}_m(x)$, $\hat{\mu}_m(x)$, and $\hat{\sigma}_m(x)$ for $1 \leq m \leq M$. 
For every chest CT scan in the LIDC-IDRI dataset, only $r \times 100\%$ of the voxel positions are used for parameter estimation. 
Different sample sizes can be examined by varying the sampling ratio $r$ in $(0, 1]$. 
For an intuitive understanding, we arbitrarily select a replicated experiment. 
In this experiment, visual depictions of 2D slices for $\hat\pi_m(x)$, $\hat\mu_m(x)$, and $\hat\sigma_m(x)$ with $r = 1$ are presented in Figure \ref{pic: parameters}(j)--(r).
We find that those estimated functions resemble the true functions very well.
For a fixed $r$ value, the experiment is randomly replicated 1000 times on a randomly selected chest CT scan in the LIDC-IDRI dataset.
% , whose seriesuid is 1.3.6.1.4.1.14519.5.2.1.6279.6001.198698492013538481395497694975.mhd
% This leads to a total of 882 RMSE values for each estimate of interest. 
The sample means are summarized in Table \ref{Table: MSE Table}. 
We find that as the value of $r$ increases, the reported mean RMSE values steadily decrease toward 0 for every estimate. 
This suggests that they should be consistent estimates of the target parameters. 
Remarkably, the convergence rate shown in Table \ref{Table: MSE Table} seems to be slow. 
This is expected because we are dealing with a nonparametric kernel-smoothing procedure in a 3D space. 
The optimal convergence rate is indeed slow.

\subsection{Case Illustration}

\begin{figure}[ht]
    \centering
    {\includegraphics[scale=0.48]{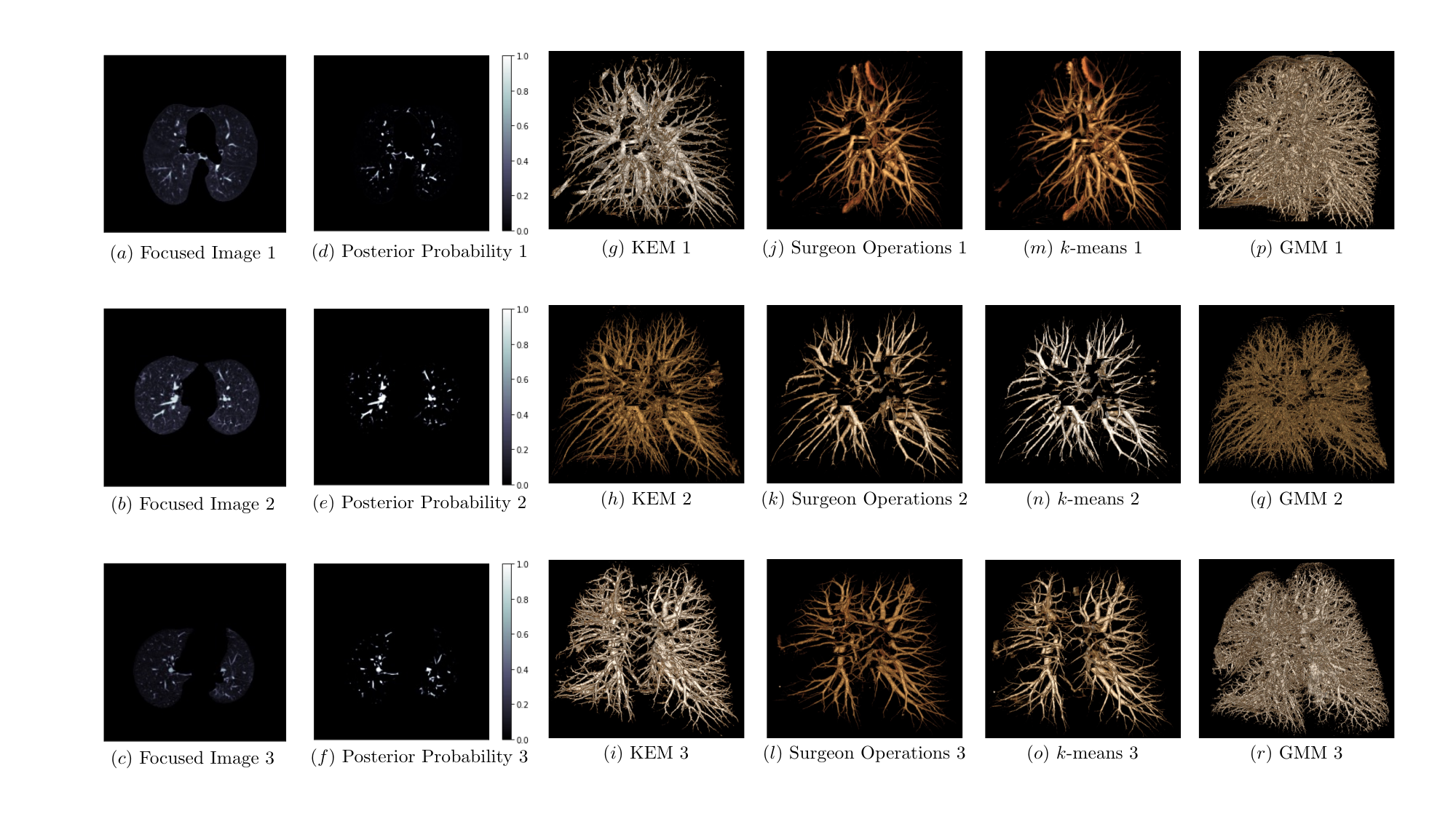}}
    \vskip -0.01in
    \caption{Graphical displays of blood vessels. 
    Panels (a)--(c) show the concentrated 2D slices of the randomly selected CTs.
    Panels (d)--(f) show 2D slices of the posterior probabilities of the randomly selected CTs. 
    Panels (g)--(i) display the 3D reconstructions of the blood vessels by the KEM algorithm. 
    Panels (j)--(l) display the 3D reconstructions of the blood vessels by a surgeon.
    Panels (m)--(o) display the 3D reconstructions of the $k$-means method.
    Panels (p)--(r) display the 3D reconstructions of the GMM method.
    }
    \label{pic: case illustration}
\end{figure}

Recall that the ultimate goal of our method is to reconstruct fine-grained 3D images of blood vessels. 
We aim to accomplish this critical task by using chest CT data in a fully automatic manner.
To this end, we randomly select three CT files from the LIDC-IDRI dataset. 
The PatientIDs of the CTs are, respectively, LIDC-IDRI-0405, LIDC-IDRI-0811, and LIDC-IDRI-0875.
% Three arbitrarily selected slices are shown in the top panels of Figure. It can be seen that a wide variety of human tissue is expressed in the CT images. Consider Figure.(c) as an example: The red triangle indicates a patient's heart. The yellow square corresponds to adipose tissue. The green circle covers some of the blood vessels, which are small light parts compared to the dark background. Therefore, blood vessels make up only a small proportion of the CT image and are easily buried in the crowd of human tissue.
To avoid distractions from undesirable human tissue, 
% we first concentrate on a more focused image that satisfies the following two requirements. First, unwanted human tissue should be excluded from the image as much as possible. Second, blood vessels should be preserved in the image.
% To this end, 
we follow \cite{lung_extraction_2019} and generate a lung mask for each CT slice.
Then, we multiply the lung mask by each slice. 
This leads to more concentrated chest CT slices, as shown in Figure \ref{pic: case illustration}(a)--(c). 
Consider Figure \ref{pic: case illustration}(c) as an example: The background is pure black.
There is a lung-shaped area floating in the middle of the image, the color of which is slightly lighter than the background. 
Blood vessels, which appear as light spots, are dispersed inside the lung-shaped area. 
Because image signals from irrelevant organ tissue are excluded, detecting blood vessels in the focused chest CT image is much easier than in the raw chest CT image.
% Remarkably, no human intervention is required for the entire image-concentrating process.

Next, we apply the developed KEM method to the concentrated chest CT data. 
%We fix $M = 3$, where one component corresponds to the background, another component represents the lung tissue, and the last component represents the blood vessels inside the lung. {Here, we would like to discussion a little more about the selection of latent classes. 
In this case, the surgeon's primary objective is to accurately identify the blood vessels.
Therefore, those voxels associated with blood vessels form one latent class. 
Second, since chest CT is mainly for diagnosing early-stage lung cancer, the lung tissue is also of great interest. 
Thus, the voxels associated with lung tissue form another important category. 
Lastly, in order to extract the lung tissue from the entire CT image, one needs to separate the background from the lung tissue. 
Therefore, the voxels associated with the background constitute the third latent class.
It seems that all of these three classes can cover almost all the voxel positions in this real example.
Therefore, we set $M=3$ in this experiment.

The CT values are then rescaled to between 0 and 1. 
We initialize $\hat{\mu}_m^{(0)}(x)$ for any $x \in \mathcal{V}$ by using a standard $k$-means algorithm.
We set $\hat{\pi}_{im}^{(0)} = \hat{\pi}_m^{(0)}(x) = 1 / M$ and $\hat{\sigma}_m^{(0)}(x) = 0.05$ for $1 \leq i \leq N$, $1 \leq m \leq M$, and $x \in \mathcal{V}$. 
The filter size is set to 3 for fast computational speed. 
Subsequently, the KEM algorithm is used to estimate the parameters of interest.
In our case, 
% class $m=1$ corresponds to blood vessels.
the estimated posterior probabilities with respect to the blood vessels are of primary interest.
% {Denote the posterior probabilities of class $m$ for all the voxels as $\lbrace \hat\pi_{im} \rbrace_{i=1}^N$, where $N$ denotes the number of voxels.}
See Figure \ref{pic: case illustration}(d)--(f) for a visual representation of the 2D slices. 
The original lung-shaped area in Figure \ref{pic: case illustration}(a)--(c) no longer obstructs the dance of the blood vessels in Figure \ref{pic: case illustration}(d)--(f).
Moreover, the blood vessels are successfully highlighted, whereas all other irrelevant human tissue is discarded. 
Subsequently, the estimated posterior probabilities with respect to the blood vessels are rescaled back to the original data range and saved as a Dicom file for the subsequent 3D reconstruction. 
We then utilize RadiAnt software to load the Dicom file and then reconstruct the blood vessels; see Figure \ref{pic: case illustration}(g)--(i) for example.
As shown, the blood vessels in the lung are preserved completely, while all the other human tissue is beautifully erased.

% It is evident that the above reconstruction procedures of the KEM method significantly differ from common practices.
% In common practice, blood vessels are directly reconstructed using the raw voxel reading $Y_i$s. 
% Although the most typical real-world reconstruction of the blood vessels is to start with the raw CT data,
Additionally, we ask the third author of this work (an experienced surgeon working in one of the major hospitals in Beijing, P.R. China) to manually reconstruct the blood vessels.
To ensure a fair basis for comparisons, the surgeon initiated the reconstruction by using the same concentrated CT as those employed in the KEM method.
% Moreover, two critical tuning parameters (WL:478 and WW:1678) must be subjectively selected by the corresponding surgeons to {touch} the blood vessels.
% Obviously, it is time-consuming, subjective, and requires substantial experience. 
% To understand the time cost required for such a subjective fine-tuning process, we consider the selected CT files, and the commonly used commercial software RadiAnt as a concrete example. 
% It took approximately ?? for the third author of this work (an experienced surgeon working in one of the major hospitals in Beijing, P.R. China) to tune the two tuning parameters for a {concentrated} CT file.
% The surgeon dedicated approximately { minutes} to complete the reconstruction task per CT. 
The reconstruction results are visualized in Figure \ref{pic: case illustration}(j)--(l).
Moreover, we used both the $k$-means and GMM methods for blood vessel reconstruction. 
The respective outcomes are visually presented in Figure \ref{pic: case illustration}(m)--(r).
First, the GMM method seems over-detailed.
Its 3D reconstruction results are distinct from those of the surgeon's manual operations.
A simulation with ground truth blood vessels generated from the result of the KEM method is given in Appendix D.
Although the ground truth blood vessels are not over-detailed, the result of the GMM method is still over-detailed.
This confirms that the excess details of the GMM method are largely due to estimation errors.
Second, the 3D reconstruction results of the KEM method, the $k$-means method, and the surgeon's operations are similar to each other.
However, the connectivity quality of blood vessels achieved through the KEM method appears superior to that of both the $k$-means method and the surgeon's operations.
This is expected because manual operations and the $k$-means method cannot adapt to subtle spatial variations within the 3D CT space.
% By nonparametrically varying the mean and variance functions, we can significantly enhance the reconstruction performances in this aspect.
Third, the KEM result is free of both the disconnected problem and the over-detailed problem; see Figure \ref{pic: case illustration}(g)--(i) for example.
This is because KEM is a more accurate estimation method due to its adaptability to spatially varying parameters in the CT space.
This makes KEM a more reliable method.

\begin{figure}
    \centering
    \includegraphics[width=0.35\linewidth]{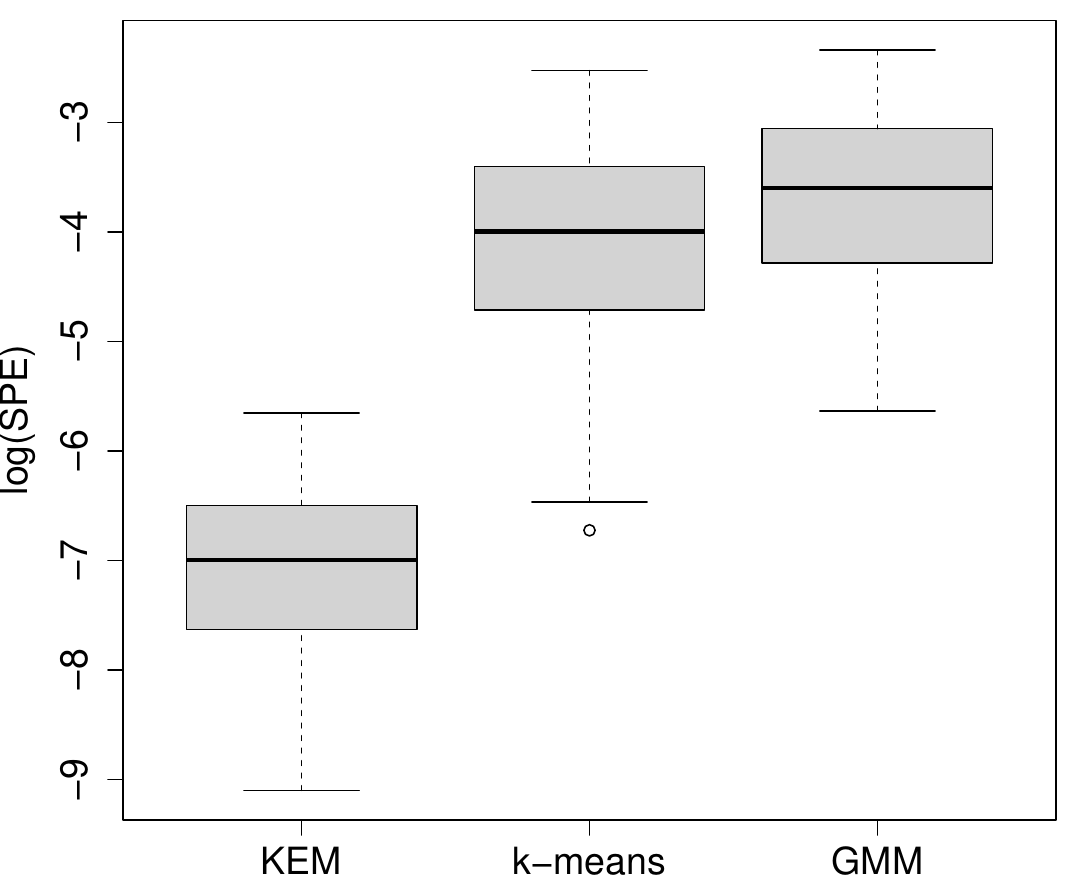}
    \vskip -0.1in
    \caption{Logarithm of the SPE results of the KEM, $k$-means, and GMM methods.}
    \label{fig: SPE}
\end{figure}

Furthermore, we also conduct a comparative analysis for prediction in terms of the SPE for the 882 CT files in the LIDC-IDRI dataset, using the KEM, $k$-means, and GMM methods on the testing set.
The SPE metric is computed as $\text{SPE} = |\mathcal{I}_1|^{-1} \sum_{i \in \mathcal{I}_1}(Y_i - \hat Y_i)^2$, where $\mathcal{I}_1$ denotes the testing set, $Y_i$ denotes the CT value, and $\hat Y_i$ denotes the predicted CT value by one particular method (i.e., the KEM, $k$-means, and GMM methods).
For the KEM method, the predicted CT value is given by $\hat Y_i = \sum_{m=1}^M \hat \pi_m(X_i) \hat \mu_m(X_i)$, for which both $\hat \pi_m(X_i)$ and $\hat \mu_m(X_i)$ are bandwidth-dependent estimates.
In contrast, for the other two competing methods (i.e., the $k$-means and GMM methods), the predicted CT value is computed by $\hat Y_i = \sum_{m=1}^M \hat \pi_m \hat \mu_m$, where $\hat \pi_m$ and $\hat \mu_m$ are the parameter estimates computed by the $k$-means method or the GMM method, which are bandwidth-independent.
Here, we use the SPE metric rather than the RMSE metric because one should know in advance the values of the true parameters to compute RMSE.
In simulation, we can compute the RMSE for each method because the true parameters are known to us.
Unfortunately, in real data analysis, the ground truth values of the parameters are unknown.
Therefore, we are unable to calculate the RMSE for the real data analysis.
In contrast, SPE remains to be computable without the true parameters.
Figure \ref{fig: SPE} illustrates boxplots of the logarithms of the SPE results. 
It is evident that the KEM method outperforms both the $k$-means and GMM methods.
In Appendix D, we provide example slices of the reconstructed responses of the three methods.
We find that the reconstructed responses of the KEM method outperform the other alternatives.

\section{CONCLUDING REMARKS}

{In summary, we developed a semiparametric Gaussian mixture model (GMM) tailored for an innovative application of 3D blood vessel reconstruction.
This work contributes significantly to both the literature on statistical theory and medical imaging applications.
From a theoretical standpoint, we introduced a semiparametric extension to the classical GMM.
We also developed a kernel-based expectation-maximization (KEM) algorithm for model estimation.
To underpin this methodology, we established a rigorous asymptotic theory.
Additionally, we devised a regression-based approach for optimal bandwidth selection, which surpasses the traditional cross-validation method in both computational and statistical efficiency.
On the application front, we addressed a critical challenge in medical imaging—3D blood vessel reconstruction—by defining it within a statistical framework of semiparametric Gaussian mixtures. 
In conclusion, we would like to discuss a topic for future research. 
The proposed method allows only a fixed number of mixture components. 
However, the human body is a highly sophisticated system with many different organs and tissues. 
Therefore, exploring the extension of the current method to accommodate a diverging number of mixture components is a topic worthy of future investigation.}

\section{Software}
\label{sec5}

Software in the form of R and Python codes, together with sample
input data and complete documentation are available online at
\url{https://github.com/Helenology/Paper_KEM}.

\section{Supplementary Material}
\label{sec6}

Supplementary material is available online at
\url{http://biostatistics.oxfordjournals.org}
% \href{http://biostatistics.oxfordjournals.org}%
% {http://biostatistics.oxfordjournals.org}.

\section*{Acknowledgments}

Jing Zhou's research is supported in part by the National
Natural Science Foundation of China (No. 72171226) and the National Statistical Science Research Project (No. 2023LD008). Ying Ji's research is supported in part by the National
Natural Science Foundation of China (No. 62306189). 
Hansheng Wang's research is partially supported by National Natural Science Foundation of China (No. 12271012).
{\it Conflict of Interest}: None declared.

\bibliographystyle{abbrvnat}
\bibliography{refs}

\newpage

% %% Figure 5
% \begin{figure}[htbp] 
%     \centering
%     \subfigure[Optimal Bandwidth Constant $C_h$]{
%     \includegraphics[scale=0.25]{pic/Ch_boxplot.pdf}}
%     \quad
%     \subfigure[Time Cost of Computation]{\includegraphics[scale=0.25] {pic/Ch_Time_boxplot.pdf}}
%     \caption{
%     }
%     \label{boxplot: optimal bandwidth boxplot}
% \end{figure}

% \begin{figure}[htbp] 
%     \centering
%     \subfigure[Optimal Bandwidth Constant $C_h$]{
%     \includegraphics[scale=0.16]{pic/Ch_boxplot.pdf}} \quad 
%     \subfigure[Logarithm of Time Duration in Seconds]{
%     \includegraphics[scale=0.16]{pic/Ch_Time_boxplot.pdf}} \quad
%     \subfigure[Testing Accuracy]{
%     \includegraphics[scale=0.16]{pic/ACC_boxplot.pdf}} \\
%     \subfigure[$\mbox{RMSE}(\hat{\pi})$]{
%     \includegraphics[scale=0.16]{pic/pi_rmse_boxplot.pdf}} \quad
%     \subfigure[$\mbox{RMSE}(\hat{\mu})$]{
%     \includegraphics[scale=0.16]{pic/mu_rmse_boxplot.pdf}} \quad
%     \subfigure[$\mbox{RMSE}(\hat{\sigma})$]{
%     \includegraphics[scale=0.16]{pic/sigma_rmse_boxplot.pdf}}
%     \caption{Panel (a) is the boxplot of the optimal bandwidth constant $C_h$ selected by the CV and REG methods. Panel (b) is the logarithm of the time duration of the two methods in seconds. 
%     Panel (c) is the boxplot of the testing accuracy of the KEM under two bandwidth selection methods, the $k$-means method, and the GMM method.
%     Panels (d)--(f) represent $\mbox{RMSE}(\hat{\pi})$, $\mbox{RMSE}(\hat{\mu})$, and $\mbox{RMSE}(\hat{\sigma})$ of the concerned methods, respectively.}
%     \label{boxplot: bwd mse bosplot}
% \end{figure}

%%% Table 1
{\small
\begin{table}[htbp]
\centering
\caption{Mean RMSE values under different sampling ratios. }\label{Table: MSE Table}
\vskip 0.1in
\begin{tabular}{@{}cccc@{}}
\toprule
$r$  & RMSE($\hat\pi$) & RMSE($\hat\mu$) & RMSE($\hat\sigma$) \\ \midrule
0.01 & 0.07398         & 0.04150         & 0.02299            \\
0.10 & 0.04552         & 0.02267         & 0.01368            \\
0.50 & 0.03031         & 0.01664         & 0.01138            \\
1.00 & 0.02440         & 0.01457         & 0.01002            \\ \bottomrule
\end{tabular}
\end{table}}

% Moreover, if the depth of the CT data is too large, the convolutional operations may not be comfortably handled by an NVIDIA Tesla P100-16GB GPU sensor with 16GB GPU memory and 64GB CPU memory. 

\end{document}